# Effect of Sr-doping of LaMnO$_3$ spacer on modulation-doped two-dimensional electron gases at oxide interfaces

Y. Z. Chen[*], Y. L. Gan, D. V. Christensen, Y. Zhang, and N. Pryds

*Department of Energy Conversion and Storage, Technical University of Denmark, Risø campus, 4000 Roskilde, Denmark*

Modulation-doped oxide two-dimensional electron gas (2DEG) formed at the LaMnO$_3$ (LMO) buffered disorderd-LaAlO$_3$/SrTiO$_3$ ($d$-LAO/LMO/STO) heterointerface, provides new opportunities for electronics as well as quantum physics. Herein, we studied the dependence of Sr-doping of La$_{1-x}$Sr$_x$MnO$_3$ (LSMO, $x$=0, 1/8, 1/3, ½ and 1) thus the filling of the Mn $e_g$ subbands as well as the LSMO polarity on the transport properties of $d$-LAO/LSMO/STO. Upon increasing the LSMO film thickness from 1 unit cell (uc) to 2 uc, a sharp metal to insulator transition of interface conduction was observed, independent of $x$. The resultant electron mobility is often higher than 1900 cm$^2$V$^{-1}$s$^{-1}$ at 2 K, which increases upon decreasing $x$. The sheet carrier density, on the other hand, is in the range of 6.9×10$^{12}$~1.8×10$^{13}$ cm$^{-2}$ (0.01~0.03 $e$/uc) and is largely independent on $x$ for all the metallic $d$-LAO/LSMO (1 uc)/STO interfaces. These results are consistent with the charge transfer induced modulation doping scheme and clarify that the polarity of the buffer layer plays a trivial role on the modulation doping. The negligible tunability of the carrier density could result from the reduction of LSMO during the deposition of disordered LAO or that the energy levels of Mn 3$d$ electrons at the interface of LSMO/STO are hardly varied even when changing the LSMO composition from LMO to SrMnO$_3$.

---

[*] Email:yunc@dtu.dk

## I. INTRODUCTION

The discovery of two-dimensional electron gases (2DEGs) at SrTiO$_3$-based heterointerfaces, provides new opportunities for electronics.[1-7] In contrast to conventional semiconductors such as Si or GaAs where the mobile carriers originate mainly from *s* and *p* orbitals, the conduction band of SrTiO$_3$ (STO) is composed of 3*d* $t_{2g}$ orbitals with strong spin, orbital and charge interactions.[7] Therefore the 2DEG of STO interfaces exhibits a large number of interesting properties, such as 2D superconductivity[8,9], magnetism[10], and metal-insulator transitions[11,12]. However, the relatively low electron mobility (~1000 cm$^2$V$^{-1}$s$^{-1}$ at 2 K) and the high sheet carrier concentration ($10^{13}$-$10^{14}$ cm$^{-2}$) of typical STO 2DEGs have hindered the applications such as demonstration of quantum Hall effects[13-15] or the achievement of sizable field effects[16].

One of the most effective approaches to boost the electron mobility is to emulate the achievements of semiconductor materials by exploiting modulation doping, where a spacer is introduced at the interface to reduce scattering by separating spatially the mobile electrons from the positively charged donors.[17] By introducing a spacer of a single unit cell (uc) LaMnO$_3$ (LMO) at the interface between disordered LAO (*d*-LAO) and STO, we have recently enhanced the electron mobility more than 20 times and lowered the carrier density of metallic interface to the order of $10^{12}$ cm$^{-2}$ with the achievement of modulation-doped oxide 2DEGs.[18] This has enabled the observation of an unconventional quantum Hall effect at the buffered oxide interface.[15] The manganite buffer layer has an empty or partially-filled subband that is lower than the conduction band of STO, therefore, it can not only separate spatially the mobile electrons from the positively charged donors but can also act as an electron sink. Additionally, the oxide modulation-doping technique provides a new opportunity to tune the carrier density thus the electron mobility of 2DEGs by control the composition of the spacer layer, such as using the Sr-doping to control the electrons filling level thus the ability to trap electrons in LSMO as illustrated in Fig. 1(b). Herein, we reported in detail the effect of Sr doping of $La_{1-x}Sr_xMnO_3$ (LSMO) ($0 \leq x \leq 1$) on the transport properties of *d*-LAO/LSMO/STO interfaces. Importantly, by extending the Sr-doping level to $x=1$, where $SrMnO_3$ (SMO) is nonpolar, we could identify whether the polar nature of the LSMO buffer layer is inherited or even enhanced in the *d*-LAO system[19]. This could help us to understand why the buffered samples show a total amount of reconstructed electrons [1.005-1.04 e per uc square area (e/uc)] much higher than that of the unbuffered sample (typically below 0.5 e/uc).

## II. EXPERIMENTAL DETAILS

The trilayer heterostructures, as illustrated in Fig. 1(a), were grown by pulsed laser deposition using singly $TiO_2$-terminated 5×5×0.5 mm (001) STO as substrates. Single-crystalline LAO and ceramic LSMO pellets ($x=0$, 1/8, 1/3, ½ and 1) were used as targets for the deposition of *d*-LAO and LSMO films, respectively. The distance between the target and the substrate was 5.5 cm. A KrF ($\lambda=248$ nm) laser with a repetition rate of 1 Hz and laser fluence of 4.0 J/cm$^2$ was adopted. During film growth, an epitaxial LSMO layer was deposited under optimized condition at 600 ºC and $1\times10^{-4}$ mbar of $O_2$, with the thickness accurately controlled on unit cell scale by monitoring *in situ* reflection high-energy electron diffraction (RHEED) intensity oscillations. Following the deposition, the LSMO/STO structure was slowly cooled to room temperature at the deposition pressure ($10^{-5}$-$10^{-4}$ mbar) before switching to the *in situ* deposition *of d*-LAO films. The final deposition of *d*-LAO film took place at a temperature below 50 ˚C and resulted in the formation of amorphous-like capping films[18,20]. For the investigated samples, the film thickness of *d*-LAO is fixed at approximately 10 nm. Electrical characterization was performed using a 4-probe Van der Pauw method with ultrasonically wire-bonded aluminum wires as electrodes.

## III. RESULTS AND DISCUSSION

The conduction of *d*-LAO/STO interface comes from the Ti $t_{2g}$ electrons,[18,20] and the LSMO/STO is found to be always highly insulating. One prominent feature of the LSMO-buffered *d*-LAO/STO is the critical dependence of the interface conduction on the LSMO film thickness. Figs.2 (a) and (b) show the room-temperature sheet conductance, $\sigma_s$, and sheet carrier density, $n_s$, respectively, of *d*-LAO/LSMO/STO heterostructures as a function of the thickness, *t*, of the LSMO buffer layer. Upon introduction of only 1 uc of LSMO buffer layer, both $\sigma_s$ and $n_s$ are decreased by more than one order in magnitude. For example, $n_s$ decreases from $1\times10^{14}$ cm$^{-2}$ for unbuffered samples to $1\times10^{13}$ cm$^{-2}$ for buffered sample. When *t* is increased to 2 uc, a sharp metal-to-insulator transition is observed. Notably, this critical thickness of *t*=2 uc for the metal-to-insulator transition is observed for all Sr doping levels ranging from *x*=0 to *x*=1, i.e. independent of the filling of the Mn $e_g$ subbands as well as the LSMO polarity.

Figures 3(a)-(c) show the temperature-dependent electrical transport properties of both unbuffered and the 1-uc-LSMO buffered *d*-LAO/STO (*d*-LAO/LSMO 1uc/STO) interfaces under different Sr doping levels. The *d*-LAO/STO interface shows a metallic behavior with electron density of approximately $1.2\times10^{14}$ cm$^{-2}$ at 300 K. Such samples often show a carrier freeze-out effect at $T\leq100$ K with the activation energy of 5-10 meV. Similar to the previous reports for *x*=0 and 1/3[18], all *d*-LAO/LSMO 1uc/STO samples exhibit a nearly temperature independent $n_s$ in addition to a more pronounced decrease in sheet resistance, $R_s$, during cooling. This suggests the suppression of localized electrons in addition to improved metallic conduction by the insertion of LSMO layers. Besides the large decrease in $n_s$, at low temperatures, all buffered samples exhibit electron mobilities, $\mu$, considerably larger than that of the unbuffered sample (~580 cm$^2$V$^{-1}$s$^{-1}$ at 2 K). Figs.4 (a) and (b) summarize the representative carrier density, $n_s$ ($T$=2 K), and the electron mobility $\mu$ ($T$=2 K), respectively, of the $d$-LAO/LSMO 1uc/STO samples as a function of $x$. Two features are noticeable: Firstly, as $x$ increases from $x$=0 to $x$=1, the $n_s$ changes negligibly. It is often suppressed in the range of 0.7-1.8×10$^{13}$ cm$^{-2}$ (0.01-0.03 electrons per uc), compared to the unbuffered sample ($n_s$=1.2×10$^{14}$ cm$^{-2}$). This is dramatically different from the case when the LSMO used as a capping layer where the sheet carrier densities monotonically decrease as increasing Sr doping.[21] Secondly, as $x$ increases from $x$=0 to $x$=1, there is a distinct decrease in the electron mobility, $\mu$, from 1.6×10$^4$ cm$^2$V$^{-1}$s$^{-1}$ to 1.9×10$^3$ cm$^2$V$^{-1}$s$^{-1}$ at $T$=2 K. However, the lower limit of electron mobility remains 3.3 time higher than the unbuffered $d$-LAO/STO.

Of central importance to understand the buffered interface is the band alignment between $d$-LAO, STO and LSMO.[18] While $d$-LAO and STO are band gap insulators, the Fermi level of LSMO is defined by the filling of $e_g$ bands.[22,23] For $x$=0, the parent compound LMO is a charge transfer insulator with electronic configuration of Mn 3$d$ $t_{2g}^3 e_g^1$, where the $e_g$ level is split into $e_{g1}$ and $e_{g2}$ levels by static Jahn-Teller distortion (Fig.1b). When $x$ increases from $x$=0 to $x$=1, The barrier height ($\Phi$) between the LSMO and the bottom of STO conduction band is increased from approximately -0.5 eV to -1.6 eV[22], as summarized in Fig. 4(c). In other words, all the LSMO buffer layers have empty or partially filled $e_g$ bands which are lower than the Ti 3$d$ $t_{2g}$ bands of STO. In this vein, the suppressed $n_s$ of the buffered samples as well as the sharp metal-to-insulator transition at $t$=2 uc can still be well explained within the charge transfer induced modulation doping scheme as reported in Ref.18: Electrons donated from the top $d$-LAO layer during the deposition first filled the empty or partially filled subband of LSMO layer ($e_g^2$ for the LMO case), and then filled the well at the interface between the STO and the LSMO, therefore achieving lower carrier density at the interface and separating spatially the mobile electrons living at the well from their positively charged donors at the top layer. However, the electronic configuration for SMO at $x=1$ is Mn $3d$ $t_{2g}^3 e_g^0$. As $x$ increases, the buffer layer in principle could trap much more electrons and the $n_s$ of the interfacial 2DEG would be further decreased. Nevertheless, we observed negligible dependence of $n_s$ over $x$, as shown in Fig. 4(a). This behavior may be understood from the following two considerations: First, the conduction of $d$-LAO/STO originates largely from oxygen vacancies on the STO side due to the redox reaction.[20] The introduction of LSMO buffer layer strongly suppresses the formation of oxygen vacancies in STO, probably because the LSMO shows negligible uptake of oxygen from STO even at high temperatures[24]. But the deposition of the $d$-LAO could get oxygen out of LSMO and reduce the stoichiometric La$_{1-x}$Sr$_x$MnO$_3$ to an oxygen deficient phase of La$_{1-x}$Sr$_x$MnO$_{2.5}$.[25] The electrons contributed by oxygen vacancies could merge the Sr-doping effect, which acts as hole-doping; Second, the electronic structure of manganite thin films is generally different from that of the bulk counterpart due to the presence of dead-layer effect[26], the negligible dependence of $n_s$ over $x$ could be due to the fact that the levels of Mn $3d$ electrons at the interface of LSMO/STO are hardly varied upon changing the LSMO composition from LMO to SMO, as reported previously[27]. Additionally, different from other doping levels, SMO is nonpolar in nature. Therefore our results further suggest that the polarity of the buffer layer has trivial effect on the modulation doping scheme and the larger amount of reconstructed electrons (~1 e/uc in LSMO and 0.005-0.04 for STO 2DEG) than 0.5 e/uc could result from the redox reduction of the LSMO during the film growth of $d$-LAO. Finally, the decrease in electron mobility upon increase $x$ could be just because of enhanced impurity scattering as the lattice mismatch of LSMO/STO become larger at higher $x$. This is supported by the fact that the highest mobility is achieved when $x \sim 1/8$, where the heterostructure shows the best lattice match.

## IV. CONCLUSION

In conclusion, we have studied the effect of Sr-doping of the LSMO buffer layer at the $d$-LAO/STO interface. Over the whole doping range of $0 \leq x \leq 1$, the introduction of a single unit cell of LSMO can suppress significantly the carrier density and increase largely the electron mobility of the interfacial 2DEG, consistent with the charge transfer induced modulation doping scheme. The highest mobility gain is achieved when $x \sim 1/8$, where LSMO shows the best lattice match with STO. The $n_s$, on the other hand, varies hardly over $x$. Furthermore, the introduction of the nonpolar SMO spacer has a trivial effect on the modulation doping scheme. The latter two effects could result from the reduction of LSMO during the $d$-LAO deposition.

**ACKNOWLEDGEMENTS**

We acknowledge the technical help from J. Geyti, and the financial support from the Denmark innovation fund.

**Figure captions:**

**Fig.1.** Schematic diagram of the charge transfer induced modulation doping at the single-unit-cell-manganite-buffered disordered-LaAlO$_3$/SrTiO$_3$. The doping of La$_{1-x}$Sr$_x$MnO$_3$ (LSMO) can, in principle, be used as a new knob to tune the carrier density of the SrTiO$_3$ 2DEG. Note that, during the formation of interface 2DEG, the electrons coming from the capping layer will firstly fill the LSMO buffer layer before the rest fill the Ti $t_{2g}$ orbitals.

**Fig.2.** (a) and (b) Room-temperature sheet conductance, $\sigma_s$, and sheet carrier density, $n_s$, respectively, of the interfacial 2DEG as a function of the thickness, $t$, of the LSMO buffer layer.

**Fig.3.** (a)-(c) The temperature dependence of sheet resistance, $R_s$, carrier density, $n_s$, and mobility, $\mu$, respectively, of $d$-LAO (10 nm)/STO heterostructures with and without LSMO ($0 \leq x \leq 1$) buffer layers.

**Fig.4.** (a) and (b) The $n_s$ ($T$=2 K) and $\mu$ ($T$=2 K) of the LSMO-buffered $d$-LAO/STO samples as a function of $x$. Note that for $x$=1/8, a representative sample of $n_s$=6.9×10$^{12}$ cm$^{-2}$, $\mu$=16205 cm$^2$V$^{-1}$s$^{-1}$, rather than the sample in Fig. 3 with the highest mobility ($n_s$=1.7×10$^{13}$ cm$^{-2}$, $\mu$=73000 cm$^2$V$^{-1}$s$^{-1}$)[18], is presented here. (c) The relative Fermi level of bulk LSMO, compared to that of STO 2DEG, $\Phi$, as a function of the LSMO doping level $x$. The insets illustrate the schematic band structure for $d$ and $p$ states of transition metals and oxygen, respectively, for STO 2DEGs, LMO and SMO.

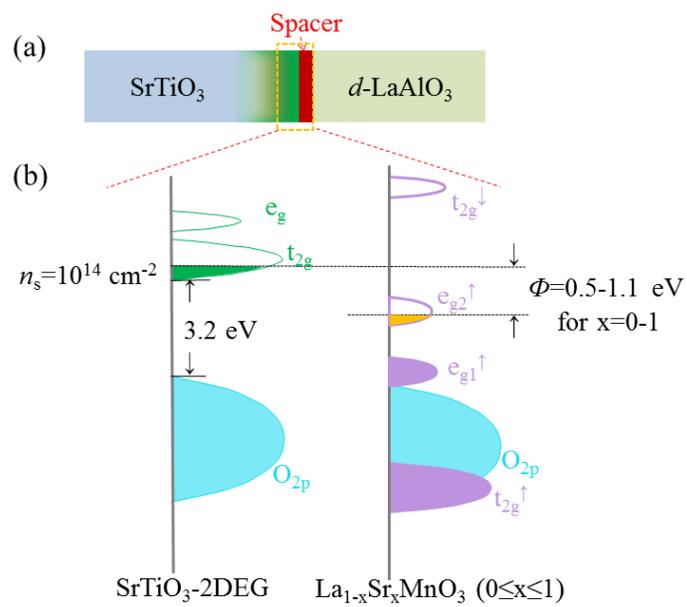

Fig. 1   *Chen et al.*

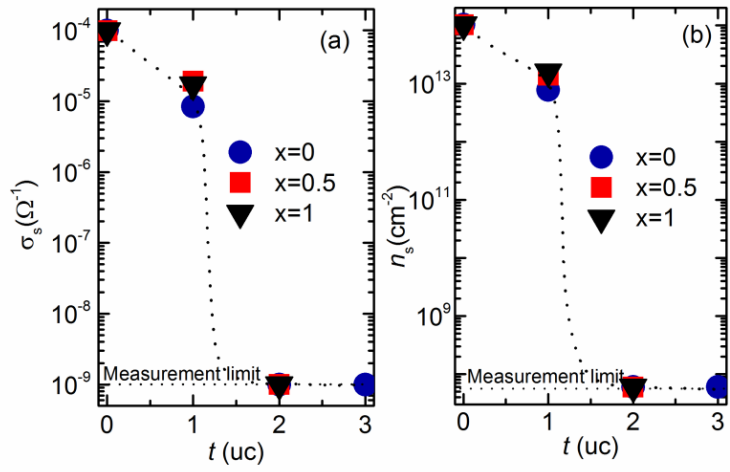

Fig.2 Chen et al.

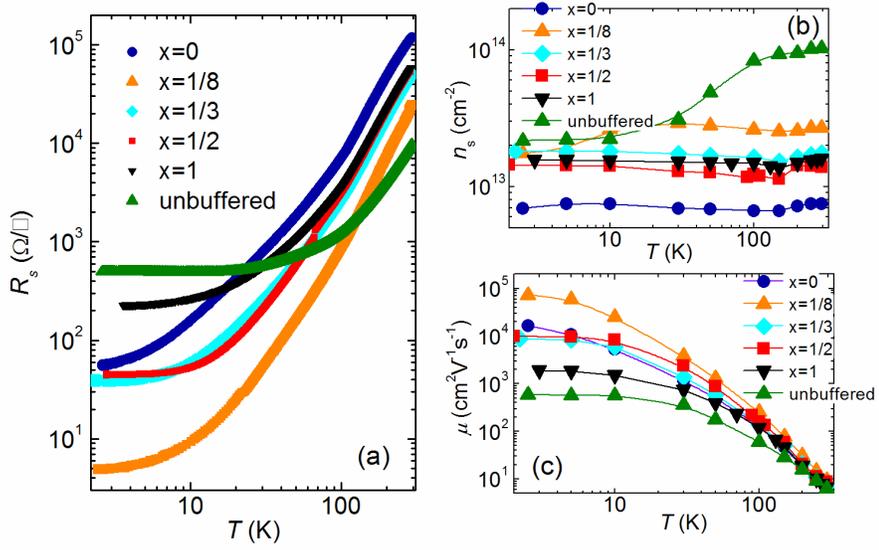

Fig.3 Chen et al.

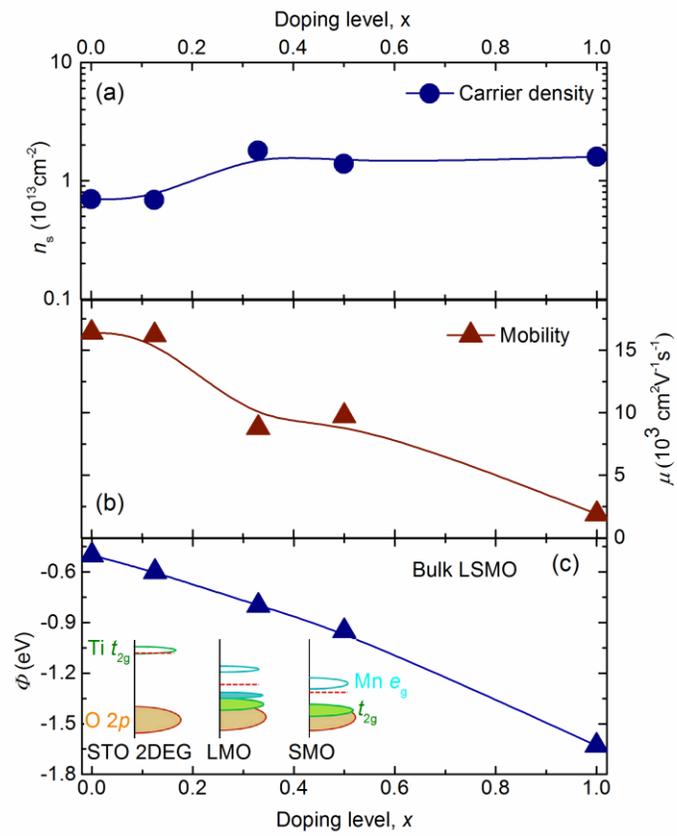

Fig. 4. *Chen et al.*